# Thermionic-enhanced near-field thermophotovoltaics

A. Datas[1,2] and R. Vaillon[3,1,4]

[1]Instituto de Energía Solar, Universidad Politécnica de Madrid, 28040 Madrid, Spain.

[2]Universitat Politècnica de Catalunya, Jordi Girona 1-3, Barcelona 08034, Spain.

[3]Univ Lyon, CNRS, INSA-Lyon, Université Claude Bernard Lyon 1, CETHIL UMR5008, F-69621, Villeurbanne, France

[4]IES, Univ. Montpellier, CNRS, F-34000 Montpellier, France

**Abstract**

Solid-state heat-to-electrical power converters are thermodynamic engines that use fundamental particles, such as electrons or photons, as working fluids. Virtually all commercially available devices are thermoelectric generators, in which electrons flow through a solid driven by a temperature difference. Thermophotovoltaics and thermionics are highly efficient alternatives relying on the direct emission of photons and electrons. However, the low energy flux carried by the emitted particles significantly limits their generated electrical power density potential. Creating nanoscale vacuum gaps between the emitter and the receiver in thermionic and thermophotovoltaic devices enables a significant enhancement of the electron and photon energy fluxes, respectively, which in turn results in an increase of the generated electrical power density. Here we propose a thermionic-enhanced near-field thermophotovoltaic device that exploits the simultaneous emission of photons and electrons through nanoscale vacuum gaps. We present the theoretical analysis of a device in which photons and electrons travel from a hot $LaB_6$-coated tungsten emitter to a closely spaced $BaF_2$-coated InGaAs photovoltaic cell. Photon tunnelling and space charge removal across the nanoscale vacuum gap produce a drastic increase in flux of electrons and photons, and subsequently, of the generated electrical power density. We show that conversion efficiencies and electrical power densities of ~ 30% and ~ 70 W/cm$^2$ are achievable at 2000 K for a practicable gap distance of 100 nm, and thus greatly enhance the performances of stand-alone near-field thermophotovoltaic devices (~10% and ~10 W/cm$^2$). A key practical advantage of this nanoscale energy conversion device is the use of grid-less cell designs, eliminating the issue of series resistance and shadowing losses, which are unavoidable in conventional near-field thermophotovoltaic devices.

## 1. Introduction

Conversion of heat into electricity is the backbone of all modern economies, generating most of world's electric power. This includes non-renewable (gas, coal, nuclear) and renewable (solar thermal) power plants. Virtually all heat engines in operation today are dynamic systems, involving generation of mechanical energy, typically a fluid flow, as an intermediate step for conversion of heat into electricity. Thermoelectric generators (TEG) [1] are a solid state alternative to dynamic systems, but their conversion efficiency is fundamentally limited by conduction heat losses, which preclude the achievement of large temperature gradients and conversion efficiencies, which are typically below 10 % [2], [3].

The search of a highly efficient alternative to TEG has been the focus of a continuous research effort since early 1950's. The two main alternatives rely on the direct emission of photons (thermophotovoltaics) [4], [5] or electrons (thermionics) [6]–[8]. In thermionics (TIC), electrons thermally emitted from a hot cathode are collected in a cold anode (or collector), and thus produce an electrical current. In thermophotovoltaics (TPV), thermally radiated photons are absorbed in a low-bandgap semiconductor and excite electron-hole pairs, which are selectively collected to produce an electrical current. Both concepts use a noncontact approach, in which the solid continuity between the hot and cold reservoirs is broken. This fully eliminates the phonon transport or heat conduction losses, which are unavoidable in TEG, and theoretically enables much higher conversion efficiencies. Conversely, the energy flux and generated power density is significantly lower due to the lower energy flux carried by radiated photons and electrons. For instance, the highest conversion efficiency of TPV devices reported so far is 24 %, but the power density (0.79 W/cm$^2$ at 1312 K [9]) is much lower than that of current state of the art TEG (today's record is 22 W/cm$^2$ at 868 K [10]). Much higher power densities, in the range of 17-25 W/cm$^2$, were experimentally demonstrated for TIC operating at 1400-1700°C, but with significantly lower conversion efficiencies in the range of 7-11% [8], [11].

Increasing power density and conversion efficiency of TPV and TIC is an important and active field of research today. Concerning TIC, most of the research focus on finding materials having a low workfunction, and device architectures that eliminates space charge (e.g. micro-spacing the cathode and collector) [6]–[8]. For TPV, the main alternatives consist of developing high quality PV cells with extended spectral response [12], or exploring novel device concepts that enable increasing the energy flux of radiative power. This can be done by using light-pipes [13] or near-field arrangements [14], the latter one having the greatest theoretical potential. Near-field TPV (nTPV) was proposed by Pan et. al. [14] in 2000 as a variation of TPV in which the emitter and the photovoltaic (PV) cell are located at nanometric distances, so that photons can tunnel through the nanoscale gap and produce a significant enhancement of radiative energy transfer and generated power density. This concept has been thoroughly assessed from the theoretical point of view during the last two decades [15]–[17]. Very recently, the concept has been proven experimentally by measuring a 40-fold enhancement of the TPV output power at a gap distances of less than 100 nm [18].

In this article we establish a thermionic-enhanced near-field thermophotovoltaic (nTiPV) device, in which both the emitter and the TPV cell are covered with low workfunction coatings (enabling the emission and collection of electrons) and separated by nanoscale vacuum gaps. The nanometric spacing provides both space-charge removal and near-field radiative enhancement, and produces a drastic increase of the flow of electrons and photons through the vacuum gap. We will show that this strategy results in a larger power density and conversion efficiency than that of conventional nTPV. Besides, we will explain why nTiPV could represent a more practical and scalable solution than nTPV to realize a highly efficient alternative to TEG.

## 2. Theory

The operational mechanism of nTiPV relies on the recently proposed concept of hybrid thermionic-photovoltaics [19], according to which photons and electrons are thermally emitted from a hot cathode/emitter towards a closely spaced TPV cell (*Figure 1*). Reducing the separation distance between the emitter and the cell to nanometric scales has two consequences: first, it produces a significant enhancement of the radiative power density, and second, it eliminates the space charge effect [6]–[8]. The latter is reflected in a reduction of the energy barriers $\phi_{EM}$ and $\phi_{CM}$ (*Figure 1*-b) that are opposing electrons' emission, which subsequently produces a drastic enhancement of the electron flux through the nanoscale gap. The emitted electrons are collected at the TPV cell surface (named collector), where they recombine with the holes that are generated within the TPV cell upon photon absorption. The photogenerated electrons are extracted from the rear-side of the TPV cell and re-injected to the emitter thorough the external wiring, creating an external current (*J*). A non-negligible lead resistance ($R_{\text{lead}}$) must be considered and carefully optimized to fulfil an existing trade-off between electrical and thermal conductivity.

In this arrangement, the output voltage is enhanced with respect to the stand-alone TPV (TIC) due to the additional voltage generated by the TIC (TPV) stage, i.e. $V_{TI}$ ($V_{PV}$), as illustrated in the band-diagram of *Figure 1*-b. On the other hand, the flux of thermionically emitted electrons must equal the external TPV's photogenerated current. Thus, if the thermionic current is lower than the photogenerated one, the TPV cell will be biased near open-circuit. The reverse condition is also true: if the thermionic current is higher than that of photogenerated one in the TPV cell, the thermionic sub-device will be biased at higher voltages than that of the maximum power point (MPP) to match the current photogenerated in the TPV cell. Therefore, in order to fully exploit the contribution of both sub-devices, it is desirable that both thermionic and photovoltaic currents coincide at their respective MPP conditions. As we will see in the discussion (section 4), this condition is not strictly necessary to produce a noticeable enhancement of power density and efficiency with respect to nTPV.

It is worth mentioning that near-field operation is unacceptable for conventional TIC, where radiative losses must be minimized. Thus, an optimal gap distance exists for conventional TIC that fulfils a trade-off between electron and photon energy fluxes [20]. On the contrary, near-field radiative enhancement is beneficial for nTiPV, where evanescent waves are effectively converted into electricity in the TPV cell.

Apart from the higher energy flux, we show that a key advantage of nTiPV is that the entire front TPV cell surface behaves as a transparent electrode, enabling a 1-D carrier transport with no lateral (2-D) current flow within the TPV cell. This avoids the use of front metal grids, typically needed in nTPV devices, and virtually eliminates the ohmic and grid-shadowing losses. The absence of front metal grids in the nTiPV design significantly simplifies the practical implementation of nanoscale gaps between the emitter and the cell, then eventually enabling the scalability of this technology.

The far-field counterpart of this device [19] is being experimentally implemented in ultra-high vacuum (UHV) conditions by using dielectric micro-spacers between the emitter and the PV cell, along with a combination of borides and fluorides as thermionic coatings [21], [22]. Thermally and electrically insulated micro-spacers that withstand large temperature gradients have already enabled the experimental demonstration of micron-gap TIC [23], [24]. Sub-micron separation distances have been also experimentally realized in the frame of near-field thermal radiation experimentations [25]–[28]. Current research efforts target the use of such nano-spacers into nTPV devices [29]. The proposed conceptual device will eventually take advantage of all these developments, which are directly transferrable to the experimental implementation of nTiPV devices.

3. Methods

In this article, we examine one possible implementation of nTiPV (*Figure 1*) in which the emitter is made of tungsten coated with a thin (10 nm) $LaB_6$ layer, which is regarded as an ideal thermionic cathode due to its low workfunction (2.5-3 eV) and high electrical conductivity [7], [8], [22]. The collector consists of an extremely thin (1-3 nm) $BaF_2$ layer, which can be directly deposited on the TPV cell front surface to provide a very low workfunction, in the range of 1-2 eV [22], and negligible absorptance. The TPV cell consist of a thin InGaAs p-n junction (0.4 µm p-region on top of a 0.7 µm n-region), whose low bandgap (0.74 eV) enables photo-generation of electron-hole pairs from absorption of thermal radiation emitted at temperatures comprised between 1400 and 2000 K (see section 1 of the Supplementary Information). The rear gold reflector turns back to the emitter the photons not absorbed in the cell, and thus contribute to boosting conversion efficiency. This specific architecture assumes that the collector layer does not modify the TPV cell band diagram in a way that holes could not diffuse towards the collector. This is a reasonable assumption provided that the high doping of the p-InGaAs layer ($8 \cdot 10^{17}$ cm$^{-3}$) will enable a high tunnelling probability (field emission) if a barrier is formed at the p-InGaAs/BaF$_2$ interface. More details on the material selection criteria are given in section 1 of the Supplementary Information.

Analysis of the nTiPV device described above requires the calculation of the total net flux of photons and electrons through the vacuum gap separating the emitter and the TPV cell. The electron flux between two infinite parallel-plane surfaces can be described by the Langmuir theory [6], which assumes one-dimensional and collision-less electron flow. For the photons, the energy flux is calculated using fluctuational electrodynamics [30] and the S-matrix method for 1D-layered media [31].

According to Langmuir theory, the flow of electrons is determined by the electrostatic potential $\psi$ in the inter-electrode gap. This potential is obtained by solving Poisson's equation $d^2\psi/dx^2 = -qn_e(x)/\varepsilon_o$ together with the Vlasov-Poisson approximation for the electron distribution function $f_e(x, v_e)$, which is used to determine the electron density $n_e(x) = \int f_e(x, v_e) dv_e$ at a position $x$ within the gap, in the nonrelativistic zero-magnetic field limit. Poisson's equation can be solved analytically by assuming a half-Maxwellian distribution for the velocities $v_e$ of thermionically emitted electrons at $x = 0$ $f_e(x = 0, v_e)$, where the electrostatic potential is maximum and therefore, there are no accelerating or decelerating fields (*Figure 1*-b). Following the previous procedure, the relation between electrostatic potential and position within the gap ($x$) can be derived [6], leading to:

$$\xi = \mp \int_0^\gamma \frac{dt}{[e^t - 1 \pm e^t \text{erf}(\sqrt{t}) \mp 2(t/\pi)^{1/2}]^{1/2}} \text{ for } \xi \lessgtr 0 \quad (1)$$

where $\xi = \frac{x}{x_o}$ is the dimensionless distance, $x_o = \left(\frac{\varepsilon_o^2 k^3}{2\pi m_e q^2}\right)^{1/4} \frac{T_E^{3/4}}{J_{TI}^{1/2}}$ is the normalization distance, $J_{TI}$ is the thermionic current density, $\varepsilon_o$ is the vacuum permittivity, $m_e$ and $q$ are the electron mass and charge, $k$ is the Boltzmann constant, and $\gamma = \frac{q\phi}{kT_E}$ is named "dimensionless motive", with $\phi = q\psi$ being the electron motive. Equation (1) can be used to determine the barriers $\phi_{EM}$ and $\phi_{CM}$ (*Figure 1*-b) that will ultimately determine the current-voltage characteristic of the thermionic stage. $\phi_{EM}$ can be obtained by combining equation (1) with the Richardson equation, which determines the current density $J_{TI}$ in the case of negligible collector back-emission as [6]:

$$J_{TI} = AT_E^2 e^{\frac{-q(\phi_E+\phi_{EM})}{kT_E}} = J_{ES} e^{\frac{-q\phi_{EM}}{kT_E}} \qquad (2)$$

where $\phi_{EM}$ is an additional barrier added to that of the emitter workfunction. This equation is valid for current densities higher than the so-called critical current $J_R$, which represents the situation at which the electrostatic potential is maximum at the collector surface, i.e. $x_C = 0$, $x_E = -d$ (see *Figure 1*-b). At this critical point, the dimensionless distance is given by $\xi = \frac{-d}{x_o(J_R)}$, so that it can be expressed as a function of $\phi_{EM}$ by making use of the Richardson equation (2) for $J_R$. The resultant expression $\xi(\phi_{EM})$ can be introduced into equation (1) by making $\gamma = \frac{q\phi_{EM}}{kT_E}$ to finally obtain $\phi_{EM}$ and $J_R$.

For current densities greater than $J_R$ ($J_{ES} > J_{TI} > J_R$), a decelerating (accelerating) field exists near the emitter (collector). This situation is named space charge mode and it is the one depicted in *Figure 1*-b. In this case, $\phi_{EM}$ can be readily calculated from the Richardson equation (2), while the calculation of $\phi_{CM}$ requires determination of the distances $x_E$ (between the emitter and the maximum of the electrostatic potential) and $x_C$ (between the collector and the maximum motive). $x_E$ is obtained by solving equation (1) for $\gamma = \frac{q\phi_{EM}}{kT_E}$ and $\xi = \frac{-x_E}{x_o}$, and the distance $x_C$ can be readily obtained as $x_C = d - x_E$. Then, $\phi_{CM}$ can be calculated by solving equation (1) for $\gamma = \frac{q\phi_{CM}}{kT_E}$ and $\xi = \frac{x_C}{x_o}$. Once both $\phi_{CM}$ and $\phi_{EM}$ are known for every value of $J_{TI}$, the internal thermionic voltage $V_{TI}$ is finally calculated as $qV_{TI} = \phi_E + \phi_{EM} - \phi_C - \phi_{CM}$ (see *Figure 1*-b). In the case of $J_{TI} < J_R$, the inter-electrode potential opposes to the electrons' flow at any position in the gap. In this situation (named retarding mode) the thermionic voltage is directly given by $V_{TI} = -\phi_C - \frac{kT_E}{q} ln\left(\frac{J_{TI}}{AT_E^2}\right)$ [6].

Following the previous procedure, the values of $V_{TI}$, $\phi_{EM}$ and $\phi_{CM}$ can be determined as a function of the current density $J_{TI}$, and the energy flux of electrons can be readily calculated [6]:

$$Q_{el} = J\frac{(\phi_{max}+2kT_E)}{e}, \text{ where } \phi_{max} = \begin{cases} \phi_E + \phi_{ME} & \text{for } J > J_R \\ \phi_C + V_{TI} & \text{for } J < J_R \end{cases} \qquad (3)$$

It must be noticed that the method described above assumes that electrons are thermionically emitted from the cathode, independently of the gap distance. When spacing between the emitter and the receiver is very small (well below ~ 10 nm), electron tunneling should be also considered. Electron tunneling was proposed for cooling in thermo-tunneling devices [32]–[34] and could be useful also for power generation in nTiPV converters.

In order to determine the nTiPV conversion efficiency, an energy balance has to be established considering all kinds of energy carriers. The main contribution to the energy balance is the heat carried by the radiated photons and electrons, but also due to heat conduction through the leads (*Figure 1*). According to the Wiedemann–Franz law [6], the minimum amount of heat lost through the leads ultimately depends on the lead electrical resistance $R_{lead}$. Assuming an average lead temperature of $(T_E + T_C)/2$, this minimum heat is given by $Q_{lead} = L\,(T_E^2 - T_C^2)/2R_{lead}$ [6], where $L = \frac{\pi^2 k^2}{3e^2}$ is the Lorentz number of the metal. Besides, half of the heat generated in the leads by Joule effect is turned back to the emitter, which can be calculated as $Q_d = S^2 J^2 R_{lead}/2$, where $S$ is the device area, equal to 1 cm² in the current study.

Finally, the efficiency of the nTiPV converter can be calculated by:

$$\eta = \frac{[SJ(V_{TI} - SJR_{lead} + V_{PV})]_{\max}}{S(Q_{el} + Q_{ph}) + Q_{lead} - Q_d} \qquad (4)$$

where $[SJ(V_{TI} - SJR_{lead} + V_{PV})]_{\max}$ is the maximum output power (in Watts) of the nTiPV converter, which is obtained at a certain voltage $V = V_{TI} - SJR_{lead} + V_{PV}$, given that the values of $V_{TI}$ and $V_{PV}$ are constrained by the current-match condition $J = J_{TI} = J_{PV}$. Notice that equation (4) neglects the heat flow from the emitter to the PV cell through the nano-spacers that will eventually separate the emitter from the PV cell. These losses are not accounted for in this analysis because they could be minimized by practical means, e.g. by using tapered spacers with very small contact area [28], [29], and therefore, they do not represent a fundamental source of losses of this concept.

The reminding variables that must be calculated are the net photon energy flux ($Q_{ph}$) and the TPV current-voltage characteristic $J_{PV} - V_{PV}$. The net photon energy flux is calculated using fluctuational electrodynamics [30] and the S-matrix method for 1D-layered media [31]. The multilayer system (*Figure 1*) is composed of 4 layers sandwiched between two semi-infinite media which are respectively made of tungsten (thermal emitter, semi-infinite), lanthanum hexaboride (LaB$_6$ cathode, 10 nm thick), vacuum gap (variable thickness *d*), barium fluoride (BaF$_2$ collector, 1 nm thick), p-doped In$_{0.53}$Ga$_{0.47}$As ($N_A$= 8 10$^{17}$ cm$^{-3}$, 0.4 µm thick), n-doped In$_{0.53}$Ga$_{0.47}$As ($N_D$= 2 10$^{17}$ cm$^{-3}$, 0.7 µm thick), gold back reflector (semi-infinite). A discussion on the selection of these particular materials and the associated property data is presented in section 1 of the Supplementary Information. Calculations are made in the angular frequency interval [7.6 10$^{13}$-7.7 10$^{15}$] rad/s (wavelength interval [0.245-24.785] µm, photon energy interval [0.050-5.068] eV). This interval covers more than 95% of the Planck radiation spectrum for the emitter temperatures (1400 to 2000 K) considered in the simulations. The complex permittivity of each material is required over the aforementioned spectral range. For In$_{0.53}$Ga$_{0.47}$As, data from [35] is used for photon energies above 0.4 eV. Below that energy, the complex permittivity is calculated using a Drude model for absorption by the free carriers and a Lorentz model for absorption by the phonons with accounting for the two-mode optical phonon behavior of III-V ternary compounds [36] (see section 2 of the Supplementary Information for details). The complex permittivity of tungsten is calculated using curve-fitting of tabulated data provided in [37]. Permittivity of lanthanum hexaboride (LaB$_6$) is taken from data reported in [38] for photon energies in the interval [2-15] eV, and in [39] for photon energies below 2 eV. For missing data for photon energy below 0.28 eV, a linear extrapolation is used. The real part of the complex refractive index of barium fluoride (BaF$_2$) is taken from a dispersion relation available in [40] for wavelengths comprised between 0.15 and 15 μm. According to [41], the extinction coefficient is null in this interval. For wavelengths larger than 15 μm, tabulated data is used [41]. Bulk properties may be inappropriate for a nanometer sized layer, but in the present case it is not a major concern since BaF$_2$ is transparent for photons energies above the bandgap of In$_{0.53}$Ga$_{0.47}$As. Finally, for gold (Au) tabulated data from experiments made on evaporated gold and reported in [42] are used for wavelengths comprised between 0.3 and 24.93 μm. For smaller wavelengths, data are taken from [43].

The S-matrix method allows calculating radiation emitted by the tungsten and lanthanum hexaboride thermally emitting layers and absorbed by any layer, as a function of the vacuum gap thickness ($d$), and over spectral intervals of interest [31]. In particular, radiation power absorbed by the receiver constituted of the collector, the p-n junction and the back reflector, ($Q_{ph}$) is computed for determining the maximum (stand-alone) nTPV conversion efficiency as $\eta_{nTPV} = P_{max}/Q_{ph}$, where $P_{max} = [V_{PV} \cdot J_{PV}]_{max}$ is the PV electrical power at the maximum power point. In order to determine the values of $J_{PV}$ and $V_{PV}$, the p- and n-doped junction layers are discretized in sub-layers in order to calculate radiation power absorbed above the bandgap energy of In$_{0.53}$Ga$_{0.47}$As through interband processes, and to infer the electron-hole pair generation rate profile, needed for solving electrical transport equations. Electrical properties of In$_{0.53}$Ga$_{0.47}$As are required for solving the minority carrier (electron and hole) diffusion equations in the frame of the low-injection approximation. Mobilities of electrons and holes are found in [44] and are respectively equal to 6000 cm$^2$ V$^{-1}$ s$^{-1}$ and 200 cm$^2$ V$^{-1}$ s$^{-1}$ for the selected doping concentrations ($N_D$= 2 10$^{17}$ cm$^{-3}$, $N_A$= 8 10$^{17}$ cm$^{-3}$). Radiative and Auger recombination coefficients are taken from [44], while the impurity recombination lifetime is derived from a curve fitting expression of experimental data as a function of doping density [45] and the Matthiessen's rule (see details in section 2.1 of the Supplementary Information). With these properties, the diffusion equations are solved as in [46]–[48] in order to derive the $J_{PV}$-$V_{PV}$ characteristic and associated parameters. In particular, the electrical power output and current density at the maximum power point ($P_{max}$, $J_{PV,max}$) are respectively used for inferring the TPV converter efficiency and the TIC converter current matching that of the TPV device.

For comparison purposes, stand-alone nTPV devices are also analyzed in this article, for which ohmic losses due to lateral current flow in the top semiconductor p-type layer must be incorporated. In this case, a lumped series resistance is included that results in a lower output voltage given by $V_{PV}|_{R_S\neq 0} = V_{PV}|_{R_S=0} - R_S J_{PV}$. The value of $R_S$ strongly depends on the size and specific layout of the TPV cell. More details on the calculation of $R_S$ are given in section 2 of the Supplementary Information.

In order to find optimal parameters leading to the maximum conversion efficiency, the Nelder-Mead algorithm [49] is used. For instance, it is used to search for the MPP voltage of both TPV and TIC converters, as well as the optimum values of $\phi_E$ and $R_{lead}$ that maximize the conversion efficiency under the constraint of current-match at MPP, i.e. $J = J_{PV}|_{max} = J_{TI}|_{max}$.

## 4. Results and discussion

*Figure 2*-a shows the current density at MPP for both TIC ($J_{TI}$) and TPV ($J_{PV}$) independent converters as a function of gap distance for the case with $T_E$=2000 K and two different values of $\phi_E$. At distances larger than ~ 1 µm, TPV operates in the far-field regime and the TPV photogenerated current density ($J_{PV}$) does not depend on distance. At smaller distances, photon tunnelling begins to dominate radiative energy transfer, leading to a drastic rise in photocurrent density. On the contrary, thermionic current is constant for distances below a certain value, at which the space charge is fully eliminated. For larger gap distances, space charge creates an additional barrier ($\phi_{EM}$) to electrons' emission that results in a decrease of current density. As a result, there are two crossing points at which both TIC and TPV sub-devices produce the same current at MPP (*Figure 2*-a): one in the near field (e.g. points A and A') and another in the far field (e.g. points B and B').

As explained in section 2, the series connection between the thermionic and photovoltaic stages in the nTiPV device (*Figure 1*) makes desirable that both $J_{TI}$ and $J_{PV}$ coincide at their respective MPP. Thus, points A (A') and B (B') represent the best-case scenario at which nTiPV fully exploits both TIC and TPV contributions.

The corresponding *J-V* curves at points A (*d* = 6.4 nm) and B (*d* = 6.9 µm) of *Figure 2*-a are shown in *Figure 2*-b and *Figure 2*-c, respectively, along with the $J_{PV}$-$V_{PV}$ and $J_{TI}$-$V_{TI}$ curves of the corresponding stand-alone TPV and TIC devices. Dashed lines represent the $J_{PV}$-$V_{PV}$ curves assumed to have a lumped series resistance ($R_S$) in the range of 2 to 10 mΩ (cell size of 1 cm²). Notice that grid-less square TPV cells of 1x1 cm² have a lumped resistance in the range of 1-10

Ω (see section 3 of the Supplementary Information). Reaching such a low range of values for $R_S$ in a stand-alone TPV converter would require relatively dense front-side metallic grids on the TPV cell. Implementing such kind of metallic grids in a near-field arrangement is particularly challenging and would bring additional shadowing and dark-current losses, not accounted for in these simulations. Therefore, the results shown for stand-alone TPV converters in this article must be regarded as the upper bound for their performance. Contrary to that, in nTiPV devices current flows in a single 1-D direction and photogenerated holes are collected from the entire TPV cell front surface. Thus, lateral conduction is fully avoided, resulting in a negligible series resistance and larger power generation capacity, as clearly shown in the *J-V* curves of *Figure 2*-b (in the near field).

The negligible ohmic losses, combined with the additional voltage generated in the thermionic stage ($V_{TI}$), produces a noticeable enhancement of the electrical power density of nTiPV. This is illustrated in *Figure 3*, which shows the electrical power density as a function of gap distance for both nTPV and nTiPV devices. The figure illustrates the relevance of using low workfunctions for the emitter in the nTiPV device to fully exploit the near-field enhancement at very small gap distances. Large workfunctions produce too low thermionic currents, lower than the photogenerated one, and the PV cell is subsequently biased near open-circuit rather than at the maximum power point. Nevertheless, nTiPV clearly outperforms nTPV for most of the gap distances, despite not fulfilling the current-match condition. Only at very large gap distances (above ~10 µm) nTiPV power decreases significantly due to the decreasing thermionic current, which is attributed to the increasing space charge. *Figure 3* shows that the power density of nTiPV can be an order of magnitude greater than that of "real" nTPV devices, assumed to have a series resistance of 10 mΩ cm$^2$, and twice as great as the one of "ideal" nTPV, assumed to have no ohmic losses.

Another important observation is that nTiPV enables larger (more practical) gap distances to produce a given electrical power. For instance, a nTiPV device operating at 2000 K with $\phi_E = 2.5$ eV, $\phi_C = 2$ eV, and $d = 100$ nm provides an electrical power density of ~ 50 W/cm$^2$ (*Figure 3*), while in the best of the cases for "ideal" nTPV, a gap distance of ~ 40 nm would be required to provide such a power level. In a more realistic scenario, where ohmic losses are not neglected in the TPV cell, an even shorter gap distance would be required. But most probably, such a high-power density would be unattainable in practice by nTPV at any gap distance.

As mentioned before, the full potential of nTiPV is achieved when both thermionic and photogenerated currents are identical at the respective MPP of both TPV and TIC sub-devices. This case is illustrated in *Figure 4*, where conversion efficiency (*Figure 4*-a) and electrical power density (*Figure 4*-b) of nTiPV and nTPV are shown as a function of gap distance for an optimized emitter workfunction, in the range of 2.5 – 3 eV, and lead resistance, in the range of 0.4 – 10 mΩ (device area of 1 cm$^2$), in order to produce the maximum efficiency and the current-match between thermionic and photogenerated currents in the near field. It is clearly observed that nTiPV generally outperforms nTPV in terms of both conversion efficiency and electrical power density. This is especially true for low collector workfunctions, resulting in a larger thermionic voltage $V_{TI}$, and when there are significant ohmic losses in the stand-alone TPV device. For instance, conversion efficiencies and electrical power densities of ~ 30% and ~ 70 W/cm$^2$ are attainable by nTiPV ($\phi_C = 1$ eV) at 2000K and a gap distance of 100 nm. On the other hand, stand-alone nTPV performs much less (efficiency and electrical power density are only ~ 10 % and ~ 10 W/cm$^2$ even though grid shadowing and dark-current losses are not considered). Therefore, another advantage of nTiPV is its potential to provide higher ratios of electrical power density to conversion efficiency than nTPV. This is clearly illustrated in *Figure 4*-c, which combines the results from *Figure 4*a-b to show conversion efficiency as a function of electrical power density. nTiPV has the potential to provide few hundreds of W/cm$^2$ at conversion efficiencies greater than 20%, whereas nTPV with realistic series resistance losses provides an order of magnitude less power density at conversion efficiencies below around 10%.

In the previous analysis (*Figure 4*), we have assumed that the emitter temperature, the cathode workfunction ($\phi_E$), and the interelectrode distance (*d*) can be precisely tuned to fulfil the current match condition at MPP. However, in most applications, emitter temperature is variable, and heat-to-power converters must operate efficiently in a broad range of temperatures. In order to evaluate the sensitivity of nTiPV to temperature variations, *Figure 5* shows the conversion efficiency (*Figure 5*-a) and the electrical power density (*Figure 5*-b) of nTiPV as a function of emitter temperature, for a constant gap distance *d*=100 nm, and collector workfunction of $\phi_C = 1.5$ eV. Solid lines for nTiPV represents the case in which $\phi_E$ and $R_{lead}$ are optimized at every temperature to fulfil current-match and provide maximum conversion efficiency. Dashed lines represent nTiPV devices with fixed values of $\phi_E$ and $R_{lead}$. The results indicate that a proper selection of $\phi_E$ and $R_{lead}$ enables a broad range of temperatures for which nTiPV clearly outperforms nTPV, despite not fulfilling the current-match condition at their respective MPPs. The only condition is that the emitter workfunction $\phi_E$ is sufficiently low for a given emitter temperature, in order to enable sufficiently large thermionic emission. If the emitter workfunction is too large, a higher emitter temperature is required in order to achieve high thermionic electron flux. However, the highest conversion efficiencies are attained for large emitter workfunctions and high temperatures, due to the larger thermionic voltage $V_{TI}$. Lower $\phi_E$ provides less conversion efficiency potential but enables a larger thermionic current that provides higher efficiencies and power densities than TPV in a broader temperature range.

## 5. Conclusions

We have established a new concept of solid-state device that combines thermionics and thermophotovoltaics in the near field to boost conversion efficiency and power density of near-field TPV converters. The fact that the entire TPV cell front surface behaves as a hole-selective contact, eliminates the necessity of a front metal grid and virtually eliminates the ohmic losses due to lateral current flows. This represents a major advantage that may enable the practical implementation and scalability of this technology. According to our results, the proposed nTiPV device may provide one of the highest ratios of conversion efficiency to power density among all the existing solid-state heat-to-electrical power converters. It is worth mentioning that the emitter and receiver properties used in this work are not optimized to ensure spectral matching, i.e. optimum emission-absorption above the bandgap of $In_{0.53}Ga_{0.47}As$ and minimum absorption by the receiver below the bandgap. This means that beyond the demonstrated benefits of hybridizing TPV and TIC in the near field, there is room for improving the performances of the TPV converter through better tuning the radiative properties of the thermionic and TPV device layers that improve spectral matching.

**Acknowledgements**

This work has been partially funded by the project AMADEUS, which has received funds from the European Union Horizon 2020 research and innovation program, FET-OPEN action, under grant agreement 737054. The sole responsibility for the content of this publication lies with the authors. It does not necessarily reflect the opinion of the European Union. Neither the REA nor the European Commission are responsible for any use that may be made of the information contained therein. A. Datas acknowledges postdoctoral fellowship support from the Spanish "Juan de la Cierva-Incorporación" program (IJCI-2015-23747). R. Vaillon is thankful to the Instituto de Energía Solar at the Universidad Politécnica de Madrid for hosting him, and acknowledges the partial funding from the French National Research Agency (ANR) under grant ANR-16-CE05-0013. Authors acknowledge Prof. Antonio Martí for reading the manuscript and providing insightful comments and improvements.

**Vitae**

**Alejandro Datas** received the Ph.D. degree from the Technical University of Madrid, Spain, in 2011. He was a visiting student at MIT (USA) in 2009, postdoctoral researcher at the Tokyo Institute of Technology (Japan) in 2012, and at the Technical University of Madrid (2013-2018). Currently, he holds a postdoctoral research position at the Technical University of Catalonia. His research focuses on solar photovoltaics, thermophotovoltaics, thermionics, and high temperature energy storage and conversion.

**Rodolphe Vaillon** received the Ph.D. degree from Poitiers University, Poitiers, France, in 1996. From October 1998 to December 2018, he was with the Centre for Energy and Thermal Sciences of Lyon, CNRS, INSA-Lyon, Université Claude Bernard Lyon 1, Villeurbanne, France. He was a visiting scholar at the University of Utah (USA) in 2016 and 2017, and a visiting Professor at the Instituto de Energía Solar (Universidad Politécnica de Madrid, Spain) in 2018. He is currently a CNRS Professor ("Directeur de Recherche") at the Institut d'Electronique et des Systèmes, Univ. Montpellier, CNRS, Montpellier, France. His research focuses on nanoscale thermal radiation, the thermal behavior of photovoltaic devices, thermophotovoltaics, and electromagnetic light scattering by complex particles.

**References**

[1]  D. M. Rowe, *Thermoelectrics Handbook: Macro to Nano*. CRC Press, 2005.
[2]  J. He and T. M. Tritt, "Advances in thermoelectric materials research: Looking back and moving forward," *Science*, vol. 357, no. 6358, Sep. 2017.
[3]  W. Liu, Q. Jie, H. S. Kim, and Z. Ren, "Current progress and future challenges in thermoelectric power generation: From materials to devices," *Acta Materialia*, vol. 87, pp. 357–376, Apr. 2015.
[4]  D. L. Chubb, *Fundamentals of thermophotovoltaic energy conversion*. Elsevier, 2007.
[5]  T. Bauer, *Thermophotovoltaics: Basic Principles and Critical Aspects of System Design*. Springer, 2011.
[6]  G. N. Hatsopoulos and E. P. Gyftopoulos, *Thermionic Energy Conversion*, 2 vols. Cambridge, Massachusetts: The MIT Press, 1979.
[7]  K. A. A. Khalid, T. J. Leong, and K. Mohamed, "Review on Thermionic Energy Converters," *IEEE Transactions on Electron Devices*, vol. 63, no. 6, pp. 2231–2241, Jun. 2016.
[8]  D. B. Go *et al.*, "Thermionic Energy Conversion in the Twenty-first Century: Advances and Opportunities for Space and Terrestrial Applications," *Front. Mech. Eng.*, vol. 3, 2017.
[9]  B. Wernsman *et al.*, "Greater than 20% radiant heat conversion efficiency of a thermophotovoltaic radiator/module system using reflective spectral control," *IEEE Transactions on Electron Devices*, vol. 51, no. 3, pp. 512–515, Mar. 2004.

[10] R. He *et al.*, "Achieving high power factor and output power density in p-type half-Heuslers Nb1-xTixFeSb," *PNAS*, vol. 113, no. 48, pp. 13576–13581, Nov. 2016.

[11] A. Datas and A. Martí, "Thermophotovoltaic energy in space applications: Review and future potential," *Solar Energy Materials and Solar Cells*, vol. 161, pp. 285–296, Mar. 2017.

[12] J. Yin and R. Paiella, "Multiple-junction quantum cascade photodetectors for thermophotovoltaic energy conversion," *Optics Express*, vol. 18, no. 2, p. 1618, Jan. 2010.

[13] D. L. Chubb, "Light Pipe Thermophotovoltaics (LTPV)," presented at the Thermophotovoltaic Generation of Electricity: TPV7, 2007, vol. 890, pp. 297–316.

[14] J. L. Pan, H. K. H. Choy, and C. G. Fonstad, "Very large radiative transfer over small distances from a black body for thermophotovoltaic applications," *IEEE Transactions on Electron Devices*, vol. 47, no. 1, pp. 241–249, Jan. 2000.

[15] S. Basu, Z. M. Zhang, and C. J. Fu, "Review of near-field thermal radiation and its application to energy conversion," *International Journal of Energy Research*, vol. 33, no. 13, pp. 1203–1232, 2009.

[16] B. Zhao, K. Chen, S. Buddhiraju, G. Bhatt, M. Lipson, and S. Fan, "High-performance near-field thermophotovoltaics for waste heat recovery," *Nano Energy*, vol. 41, pp. 344–350, Nov. 2017.

[17] E. Tervo, E. Bagherisereshki, and Z. Zhang, "Near-field radiative thermoelectric energy converters: a review," *Front. Energy*, vol. 12, no. 1, pp. 5–21, Mar. 2018.

[18] A. Fiorino, L. Zhu, D. Thompson, R. Mittapally, P. Reddy, and E. Meyhofer, "Nanogap near-field thermophotovoltaics," *Nature Nanotechnology*, p. 1, Jun. 2018.

[19] A. Datas, "Hybrid thermionic-photovoltaic converter," *Applied Physics Letters*, vol. 108, no. 14, p. 143503, Apr. 2016.

[20] J.-H. Lee, I. Bargatin, N. A. Melosh, and R. T. Howe, "Optimal emitter-collector gap for thermionic energy converters," *Applied Physics Letters*, vol. 100, no. 17, p. 173904, Apr. 2012.

[21] A. Datas *et al.*, "AMADEUS: Next generation materials and solid state devices for ultra high temperature energy storage and conversion," in *AIP Conference Proceedings*, 2018, vol. 2033, p. 170004.

[22] A. Mezzi *et al.*, "Investigation of work function and chemical composition of thin films of borides and nitrides," *Surface and Interface Analysis*, vol. 50, no. 11, 2018.

[23] J. H. Lee *et al.*, "Microfabricated Thermally Isolated Low Work-Function Emitter," *Journal of Microelectromechanical Systems*, vol. 23, no. 5, pp. 1182–1187, Oct. 2014.

[24] R. Y. Belbachir, Z. An, and T. Ono, "Thermal investigation of a micro-gap thermionic power generator," *J. Micromech. Microeng.*, vol. 24, no. 8, p. 085009, 2014.

[25] K. Ito, K. Nishikawa, A. Miura, H. Toshiyoshi, and H. Iizuka, "Dynamic Modulation of Radiative Heat Transfer beyond the Blackbody Limit," *Nano Lett.*, vol. 17, no. 7, pp. 4347–4353, Jul. 2017.

[26] J. I. Watjen, B. Zhao, and Z. M. Zhang, "Near-field radiative heat transfer between doped-Si parallel plates separated by a spacing down to 200 nm," *Appl. Phys. Lett.*, vol. 109, no. 20, p. 203112, Nov. 2016.

[27] M. Ghashami, H. Geng, T. Kim, N. Iacopino, S. K. Cho, and K. Park, "Precision Measurement of Phonon-Polaritonic Near-Field Energy Transfer between

Macroscale Planar Structures Under Large Thermal Gradients," *Phys. Rev. Lett.*, vol. 120, no. 17, p. 175901, Apr. 2018.

[28] K. Ito, A. Miura, H. Iizuka, and H. Toshiyoshi, "Parallel-plate submicron gap formed by micromachined low-density pillars for near-field radiative heat transfer," *Applied Physics Letters*, vol. 106, no. 8, p. 083504, Feb. 2015.

[29] J. DeSutter, L. Tang, and M. Francoeur, "Near-field radiative heat transfer devices," Nov. 2018.

[30] S. M. Rytov, Y. A. Kravtsov, and V. I. Tatarskii, *Principles of Statistical Radiophysics 3: Elements of Random Fields*. Berlin Heidelberg: Springer-Verlag, 1989.

[31] M. Francoeur, M. Pinar Mengüç, and R. Vaillon, "Solution of near-field thermal radiation in one-dimensional layered media using dyadic Green's functions and the scattering matrix method," *Journal of Quantitative Spectroscopy and Radiative Transfer*, vol. 110, no. 18, pp. 2002–2018, Dec. 2009.

[32] Y. Hishinuma, T. H. Geballe, B. Y. Moyzhes, and T. W. Kenny, "Refrigeration by combined tunneling and thermionic emission in vacuum: Use of nanometer scale design," *Appl. Phys. Lett.*, vol. 78, no. 17, pp. 2572–2574, Apr. 2001.

[33] E. T. Enikov and T. Makansi, "Analysis of nanometer vacuum gap formation in thermo-tunneling devices," *Nanotechnology*, vol. 19, no. 7, p. 075703, 2008.

[34] Y. Hishinuma, T. H. Geballe, B. Y. Moyzhes, and T. W. Kenny, "Measurements of cooling by room-temperature thermionic emission across a nanometer gap," *Journal of Applied Physics*, vol. 94, no. 7, pp. 4690–4696, Sep. 2003.

[35] M. Muñoz *et al.*, "Optical constants of In0.53Ga0.47As/InP: Experiment and modeling," *Journal of Applied Physics*, vol. 92, no. 10, pp. 5878–5885, Oct. 2002.

[36] S. Adachi, "Chapter 31: Ternary and quaternary compounds," in *Springer Handbook of Electronic and Photonic Materials*, Springer, 2006, pp. 735–752.

[37] E. D. Palik, *Handbook of Optical Constants of Solids*, vol. 1. San Diego, 1998.

[38] Y. Sato, M. Terauchi, M. Mukai, T. Kaneyama, and K. Adachi, "High energy-resolution electron energy-loss spectroscopy study of the dielectric properties of bulk and nanoparticle LaB6 in the near-infrared region," *Ultramicroscopy*, vol. 111, no. 8, pp. 1381–1387, Jul. 2011.

[39] L. Xiao *et al.*, "Origins of high visible light transparency and solar heat-shielding performance in LaB6," *Appl. Phys. Lett.*, vol. 101, no. 4, p. 041913, Jul. 2012.

[40] H. H. Li, "Refractive index of alkaline earth halides and its wavelength and temperature derivatives," *Journal of Physical and Chemical Reference Data*, vol. 9, no. 1, pp. 161–290, Jan. 1980.

[41] M. Querry, "Optical Constants of Minerals and Other Materials from the Millimeter to the Ultraviolet," CRDEC-CR-88009, Nov. 1987.

[42] R. L. Olmon *et al.*, "Optical dielectric function of gold," *Phys. Rev. B*, vol. 86, no. 23, p. 235147, Dec. 2012.

[43] S. Babar and J. H. Weaver, "Optical constants of Cu, Ag, and Au revisited," *Appl. Opt., AO*, vol. 54, no. 3, pp. 477–481, Jan. 2015.

[44] M. E. Levinshtein, *Handbook series on semiconductor parameters, Vol. 1 Si, Ge, C (Diamond), GaAs, GaP, GaSb, InAs, InP, InSb*. World Scientific, 1996.

[45] R. K. Ahrenkiel, R. Ellingson, S. Johnston, and M. Wanlass, "Recombination lifetime of In0.53Ga0.47As as a function of doping density," *Appl. Phys. Lett.*, vol. 72, no. 26, pp. 3470–3472, Jun. 1998.

[46] M. Francoeur, R. Vaillon, and M. P. Mengüç, "Thermal Impacts on the Performance of Nanoscale-Gap Thermophotovoltaic Power Generators," *IEEE Transactions on Energy Conversion*, vol. 26, no. 2, pp. 686–698, Jun. 2011.

[47] M. P. Bernardi, O. Dupré, E. Blandre, P.-O. Chapuis, R. Vaillon, and M. Francoeur, "Impacts of propagating, frustrated and surface modes on radiative, electrical and thermal losses in nanoscale-gap thermophotovoltaic power generators," *Sci Rep*, vol. 5, Jun. 2015.

[48] E. Blandre, P.-O. Chapuis, and R. Vaillon, "High-injection effects in near-field thermophotovoltaic devices," *Scientific Reports*, vol. 7, no. 1, p. 15860, Nov. 2017.

[49] J. Lagarias, J. Reeds, M. Wright, and P. Wright, "Convergence properties of the Nelder-Mead simplex method in low dimensions," *Siam Journal on Optimization*, vol. 9, no. 1, pp. 112–147, Dec. 1998.

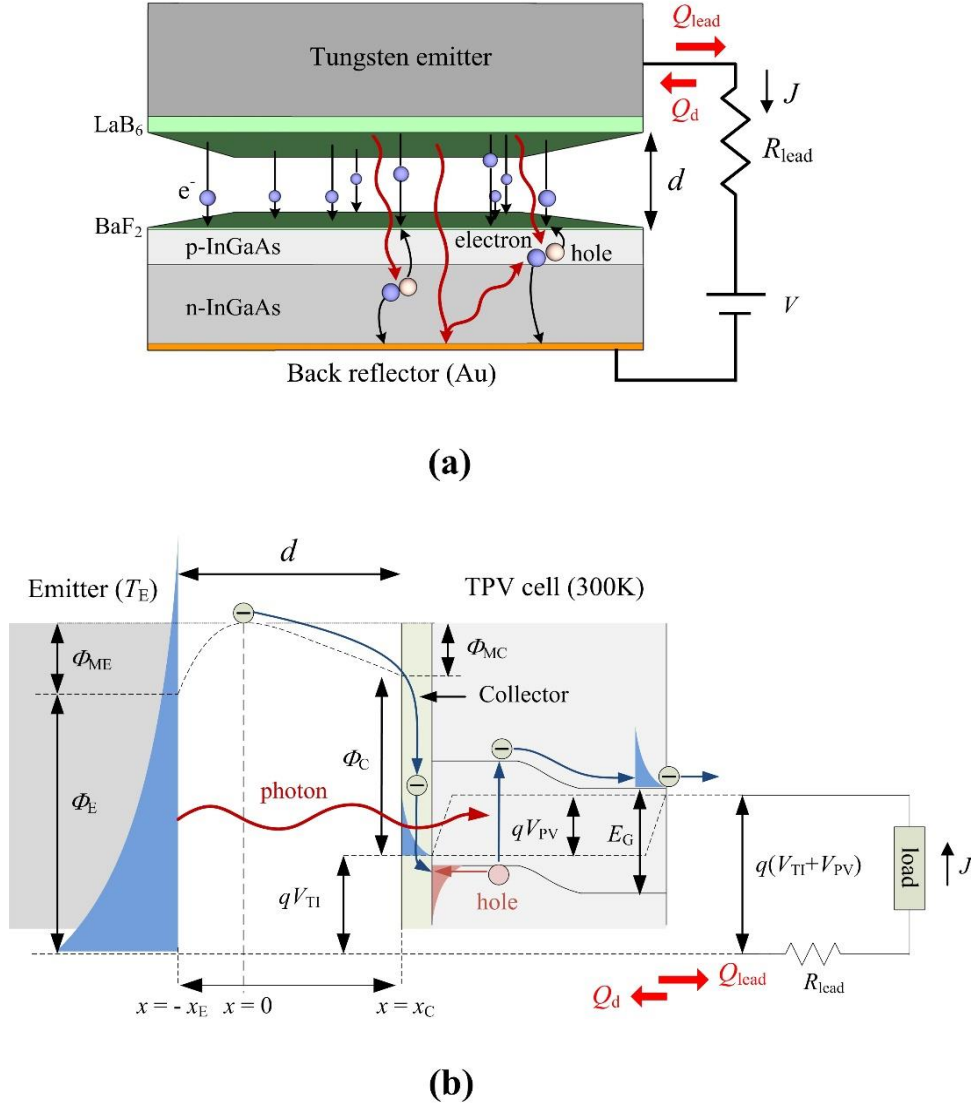

*Figure 1. (a) Sketch of the (n)TiPV converter comprising an emitter (tungsten coated with LaB$_6$) and a receiver consisting of a TPV cell (InGaAs p-n junction) covered by a thin BaF$_2$ collector. (b) Band diagram of a (n)TiPV device, where a cathode/emitter with workfunction $\boldsymbol{\phi_E}$ radiates electrons and photons towards a TPV cell separated by a distance $\boldsymbol{d}$ from the emitter and coated by a very thin transparent collector with workfunction $\boldsymbol{\phi_C} < \boldsymbol{\phi_E}$. The space charge effect within the vacuum gap creates additional potential barriers $\boldsymbol{\phi_{ME}}$ and $\boldsymbol{\phi_{MC}}$, which depend on the gap distance $\boldsymbol{d}$, and that oppose to the electron flow. Photogenerated holes in the TPV cell recombine with electrons from the collector, whereas photogenerated electrons, which have an additional chemical energy $\boldsymbol{qV_{PV}}$, diffuse to the TPV cell rear contact (n-type) and produce useful work. A lead resistance $R_{lead}$ is included to consider the resistance of the leads required for connecting the cathode/emitter. This resistance must fulfil a trade-off between thermal and electrical conductivity.*

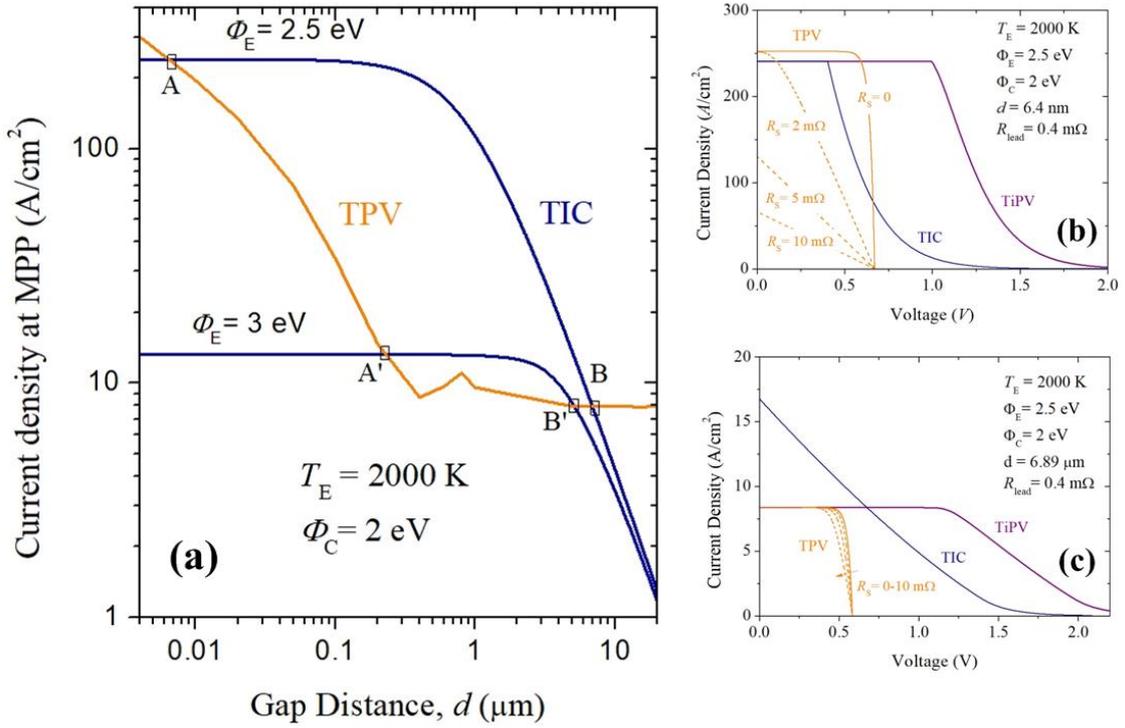

*Figure 2. (a) Current density at the MPP for stand-alone TPV (with $R_s = 0$) and TIC sub-devices as a function of the gap distance d. (b-c) Current density – Voltage (J-V) curves for the TiPV devices at current-match condition in the near field (point A) and far field (point B). The J-V curves of independent TPV and TIC sub-devices are superimposed, illustrating the series resistance effect of stand-alone TPV cells (dashed J-V curves). The lead resistance $R_{lead}$ is optimized independently for the two cases ($\phi_E = 2.5$ eV and $\phi_E = 3$ eV), in order to provide the highest conversion efficiency at the current-match condition in the near field. The device area is 1 $cm^2$.*

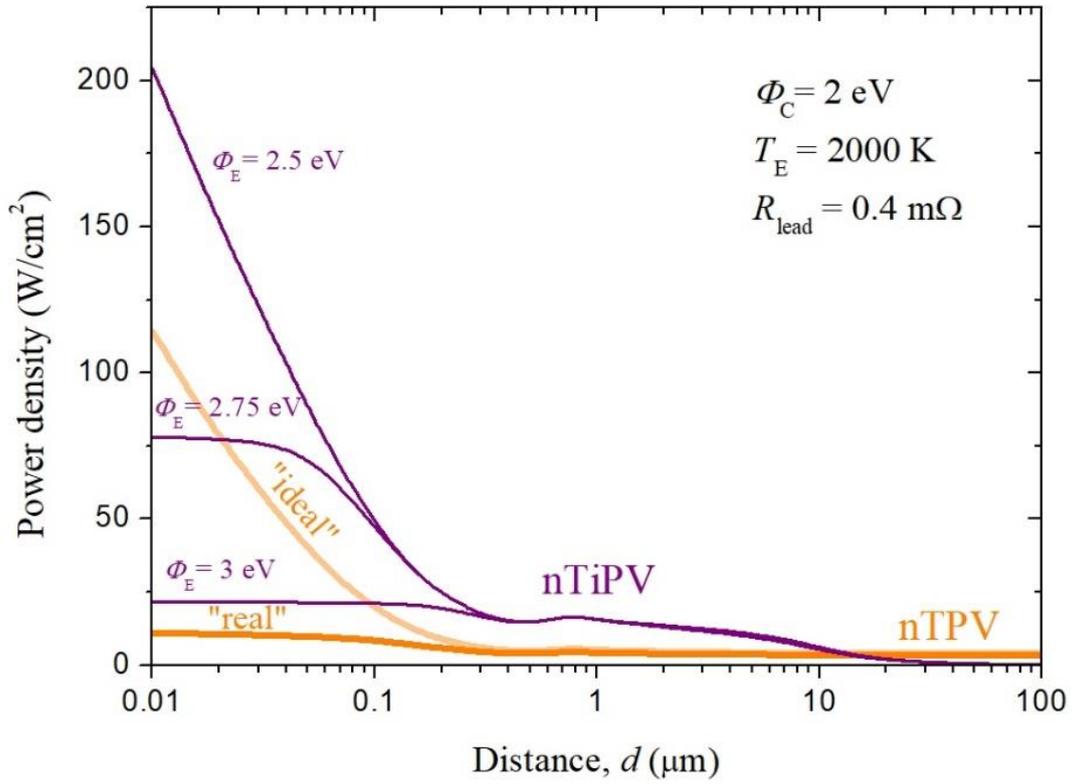

*Figure 3. Electrical power density of nTiPV and nTPV converters as a function of gap distance between the emitter and the PV cell. Different workfunctions of the emitter ($\phi_E$) are considered for nTiPV. "ideal" and "real" nTPV refers to the case with negligible ohmic losses and the more realistic case with a series resistance of 10 mΩ, respectively. The device area is 1 cm$^2$.*

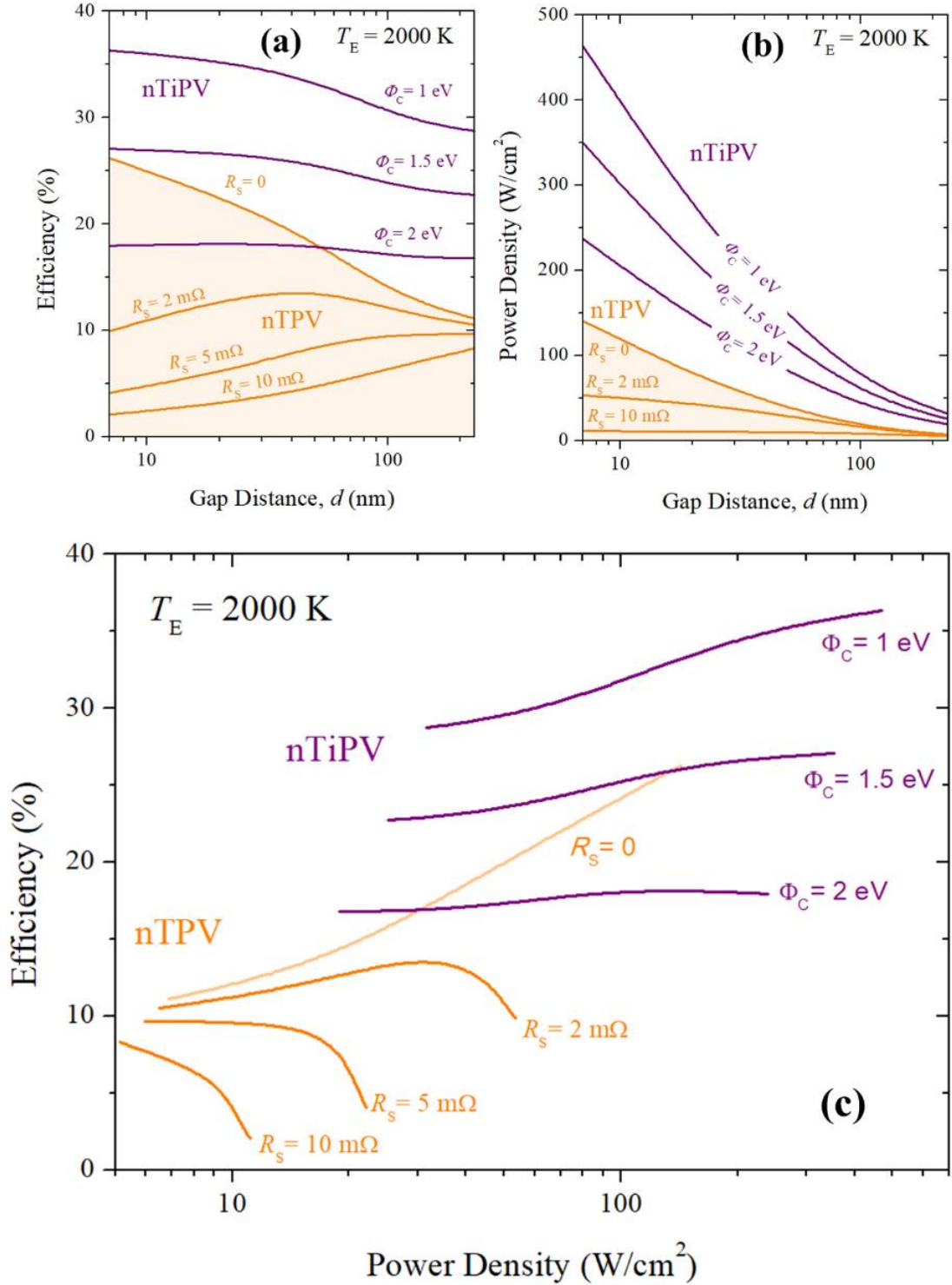

*Figure 4. (a) Conversion efficiency and (b) electrical power density of nTiPV and nTPV as a function of gap distance, (c) conversion efficiency as a function of electrical power density for nTiPV and nTPV converters, rearranged from the results shown in (a) and (b). Results are obtained for an emitter temperature of 2000 K. Both the emitter workfunction $\phi_E$ and the lead resistance $R_{lead}$ are optimized. Results for standalone TPV device with different values of series resistance are superimposed for direct comparison. The device area is 1 cm$^2$.*

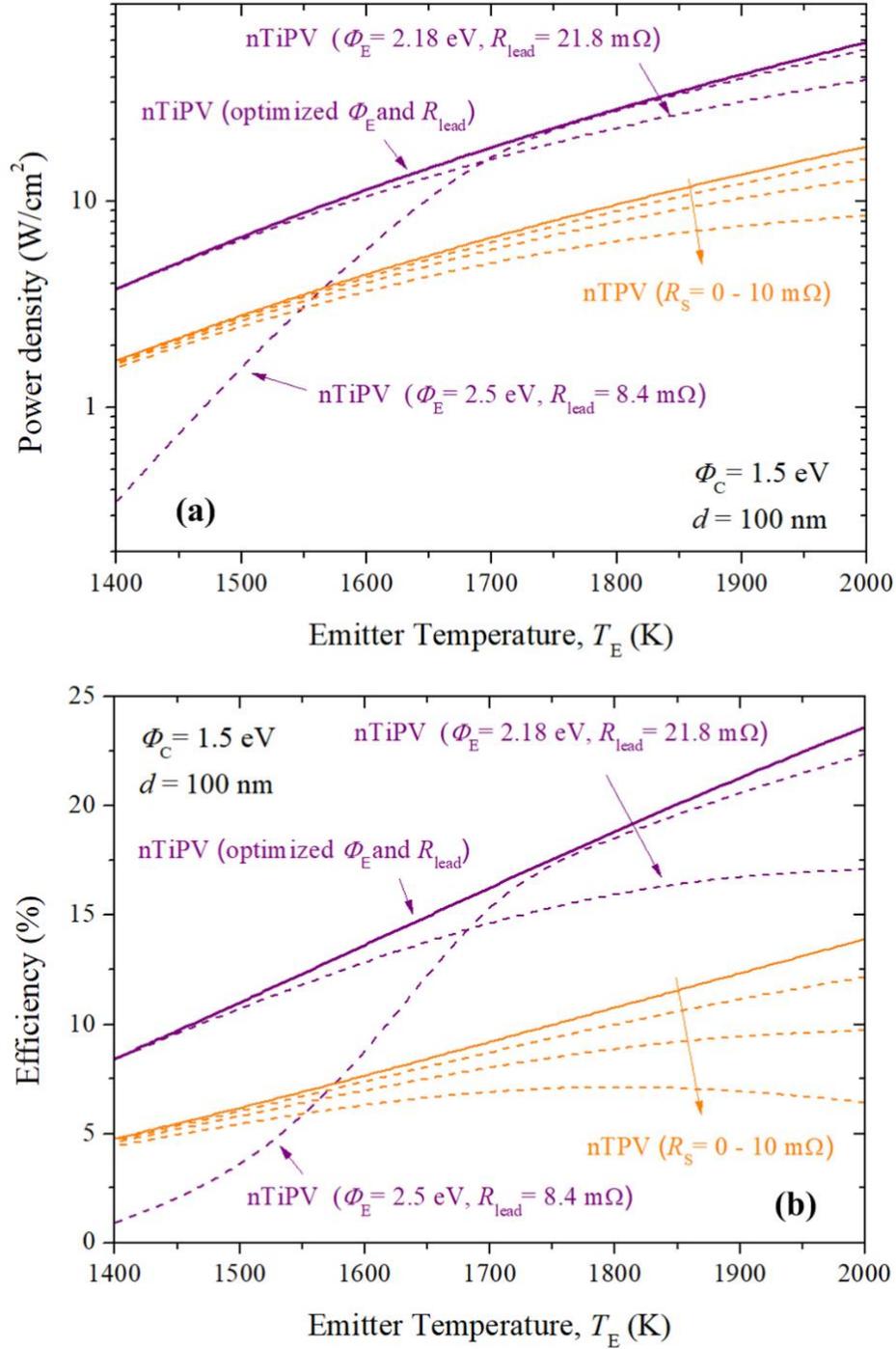

*Figure 5. nTiPV and nTPV conversion efficiency (a) and electrical power density (b) as a function of emitter temperature for a fixed gap distance (d) and collector workfunction ($\phi_C$). The solid line for nTiPV represents the case where the emitter workfunction ($\phi_E$) and lead resistance ($R_{lead}$) are optimized to maximize conversion efficiency. Dashed lines for nTiPV represent two specific choices for $\phi_E$ and $R_{lead}$. The solid line for TPV represents the cases without ohmic losses. Dashed lines for TPV represent different values of series resistance (2, 5 and 10 m$\Omega$). The device area is 1 cm$^2$.*

**Vitae (requested pictures)**

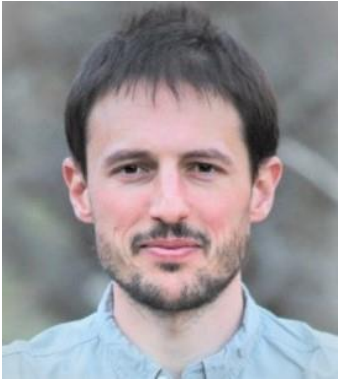

**Alejandro Datas**

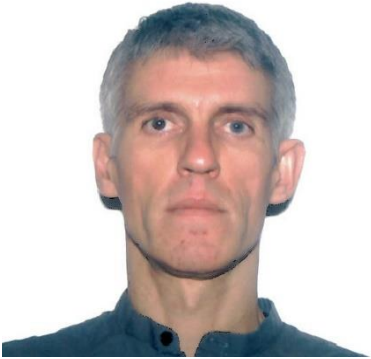

**Rodolphe Vaillon**